\documentclass[twocolumn]{aastex61}
\pdfoutput=1 
\usepackage{hyperref}
\usepackage{amsmath,amstext}
\usepackage[T1]{fontenc}
\usepackage{apjfonts} 
\usepackage[figure,figure*]{hypcap}
\newcommand{\gsim}{\lower.7ex\hbox{$\;\stackrel{\textstyle>}{\sim}\;$}}
\newcommand{\lsim}{\lower.7ex\hbox{$\;\stackrel{\textstyle<}{\sim}\;$}}

\defcitealias{HL17}{HL17}

\newcommand{\figref}[1]{Figure \ref{#1}}

\newcommand{\TSS}{\emph{TESS}}
\newcommand{\KEP}{\emph{Kepler}}

\newcommand{\NSN}{7}
\newcommand{\NPN}{14}
\newcommand{\NSE}{10}
\newcommand{\NPE}{25}

\newcommand{\NSNCons}{4}
\newcommand{\NPNCons}{6}
\newcommand{\NSECons}{6}
\newcommand{\NPECons}{12}

\newcommand{\NPNRange}{6 - 14}
\newcommand{\NPERange}{12 - 25}

\shorttitle{TTV Masses Using Kepler and TESS}
\shortauthors{Goldberg et al.}

\submitjournal{ApJ}

\begin{document}

\title{Prospects for Refining Kepler TTV Masses using TESS Observations}

\author[0000-0003-3868-3663]{Max~Goldberg}
\affiliation{Department of Astronomy and Astrophysics, University of Chicago, 5640 S Ellis Ave, Chicago, IL 60637, USA}

\author[0000-0002-1032-0783]{Sam~Hadden}
\affiliation{Harvard-Smithsonian Center for Astrophysics, 60 Garden St., MS 51, Cambridge, MA 02138, USA}
\correspondingauthor{Sam~Hadden \& Max~Goldberg}
\email{samuel.hadden@cfa.harvard.edu; maxgoldberg@uchicago.edu}

\author[0000-0001-5133-6303]{Matthew~J.~Payne} 
\affiliation{Harvard-Smithsonian Center for Astrophysics, 60 Garden St., MS 51, Cambridge, MA 02138, USA}

\author[0000-0002-1139-4880]{Matthew~J.~Holman}
\affiliation{Harvard-Smithsonian Center for Astrophysics, 60 Garden St., MS 51, Cambridge, MA 02138, USA}


\begin{abstract}
In this paper we investigate systems previously identified to exhibit transit timing variations (TTVs) in \KEP~data, with the goal of predicting the expected improvements to the mass and eccentricity constraints that will arise from combining \KEP~data with future data from the \TSS~mission.
We advocate for the use of the Kullback-Leibler (KL) divergence as a means to quantify improvements in the measured constraints.
Compared to the original \KEP~data, the \TSS~data will have a lower signal-to-noise ratio, rendering some of the planetary transits undetectable, and lowering the accuracy with which the transit mid-time can be estimated. 
Despite these difficulties, out of the 55 systems (containing $143$ planets) investigated, we predict that the collection of short-cadence data by \TSS~will be of significant value (i.e. it will improve the mass uncertainty such that the KL divergence is $\gsim0.1$) for approximately \NPNRange{} planets during the nominal mission, with the range primarily driven by the uncertain precision with which transit mid-times will be recovered from \TSS~data.
In an extended mission this would increase to a total of approximately \NPERange{} planets.
\end{abstract}

%

\section{Introduction}
\label{SECN:INTRO}
    Transit timing variations (TTVs) are a powerful tool for measuring masses and eccentricities in multi-transiting systems \citep{Agol05,Holman05,Holman10}. The \KEP~mission's \citep{Borucki2010} 4 years of nearly-continuous photometric observations provided a rich data set containing hundreds of TTV measurements for transiting planets \citep{Rowe2015,Holczer2016,Ofir2018}. This data set has yielded a significant number of mass and eccentricity measurements in  systems of small sub-Jovian planets that would otherwise be largely inaccessible to radial velocity characterization \citep[e.g.,][]{Jontof-Hutter2016,HL17}. 

    The \TSS~mission \citep{TESS15}, successfully launched on 18 April 2018, represents the next generation in  space based transit missions. \TSS~will revisit the \KEP~field in its second year of operation and the additional transit measurements of \KEP~systems obtained by \TSS~during this time will provide an opportunity to improve planet mass and eccentricity constraints derived from \KEP~transit timing data. If \TSS~operates beyond its nominal mission, it could re-visit the \KEP~field multiple times and further improve the precision TTV-derived constraints. 
    
    The goal of this paper is to quantify the expected improvements to the mass constraints\footnote{TTV parameter inference generally exhibits strong correlations between planet masses and eccentricities \citep[e.g.,][]{Lithwick12}. Therefore, we do not investigate improvements to eccentricity constraints separately since we do not expect situations in which eccentricity constraints are improved without corresponding improvements to planet mass constraints.} 
    that we predict will arise from combining \TSS~data with \KEP~data, during both the nominal \TSS~mission as well as under various extended mission scenarios.

    The systems we use in this work were previously observed to exhibit TTVs in \KEP~data.
    Specifically, we work with the 55 systems fit by \citet{HL17} (hereafter HL17). This sample of multi-transiting \KEP~TTV systems was initially selected from the \citet{Holczer2016} catalog on the significance of their TTV signals. HL17 fit the transit times of these systems derived by \citet{Rowe2015} using MCMC simulations to generate posterior samples of planets' orbital elements and planet-to-star mass ratios.  In this work, we take HL17's computed posterior samples of orbital elements and planet-to-star mass ratios as the starting point of our analysis. \citet{Rowe2015} derived transit times from \KEP~long-cadence data so the \KEP~transit mid-time uncertainties discussed below are derived from 30-minute exposures.

In Section \ref{s:TESS} we describe our expectations for the \TSS~data; 
In Section \ref{s:Methods} we describe our statistical methods; 
In Section \ref{s:Nominal} we describe the systems likely to exhibit improved mass measurements under both \emph{nominal} and \emph{extended} \TSS~mission scenarios, 
and finally in Section \ref{SECN:DISC} we discuss our results and conclusions.


\section{Expected Observational Uncertainties}
\label{s:TESS}

We computed the \TSS~magnitudes of the 55 multiplanet system hosts with the \texttt{ticgen} python module \citep{ticgen,TIC2017}, using J, H, and Ks magnitudes taken from the Exoplanet Archive. 
For each star, we estimated the photometric uncertainty in the \KEP~data using \KEP~magnitudes from the Exoplanet Archive and pre-launch noise characteristics from Kepler Science Center\footnote{\url{https://keplergo.arc.nasa.gov/CalibrationSN.shtml}}.
We estimated the photometric uncertainty in a single \TSS~short cadence measurement using the computed \TSS~magnitude and the expected noise characteristics from \cite{TESS15}. 

\subsection{\TSS~Transit-Time Uncertainties}
\label{s:Uncertainty}
\begin{figure*}[thp]
        \centering
        \includegraphics[width=0.95\textwidth]{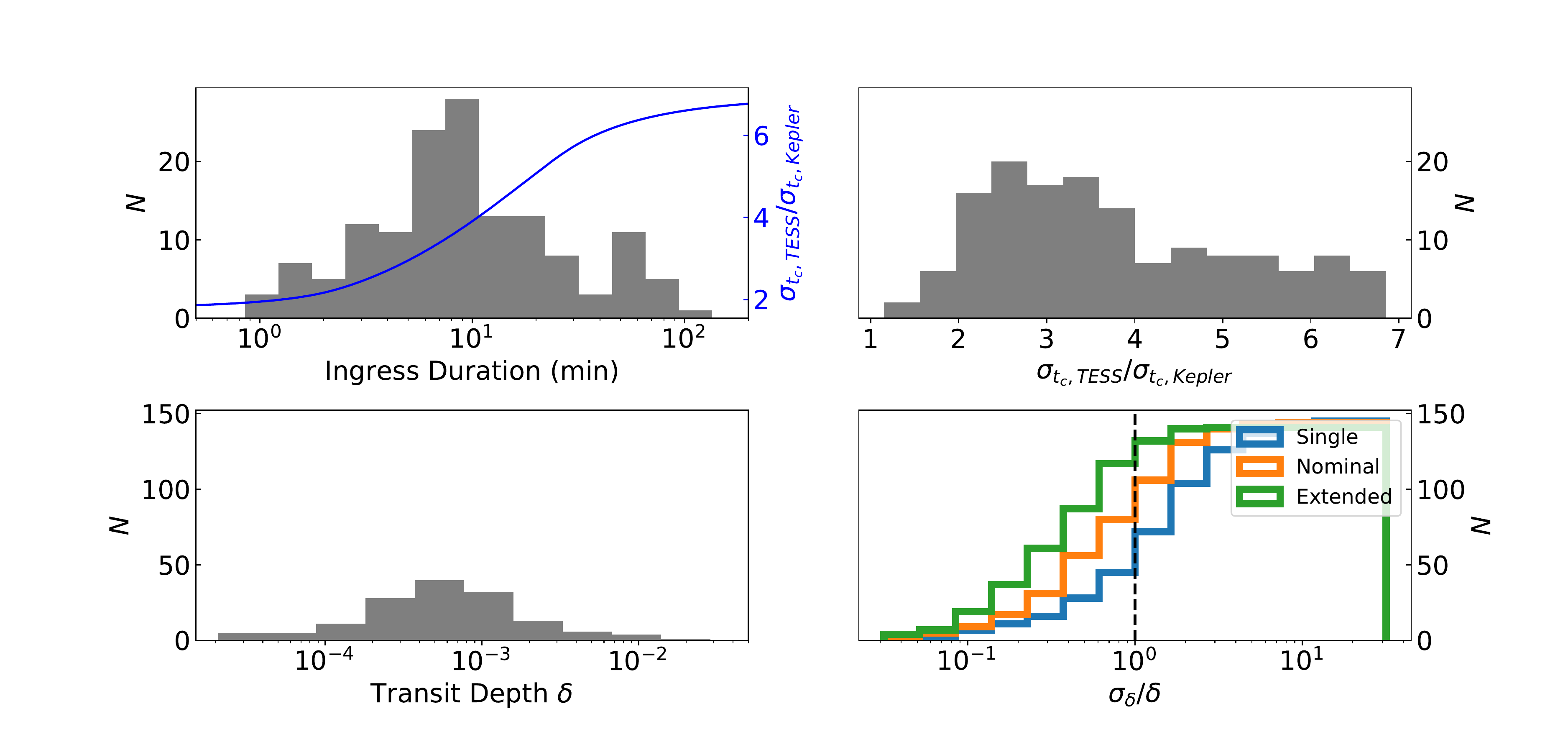}
        \caption{
        {\bf Top left:} Grey bars: distribution of ingress duration for 143 TTV planets. Blue line: relative uncertainty in transit mid-time as a function of ingress duration, for an example star with a \KEP~magnitude of 10.6 and \TSS~magnitude of 10.0.
        {\bf Top right:} Distribution of ratios of uncertainties in transit mid-times between \KEP~(long cadence) and \TSS~(short cadence) observations for the 143 \KEP~TTV planets. 
        For the systems under consideration, the uncertainty in the transit time from \TSS~is typically a few times worse than from the \KEP~data.
        {\bf Bottom left:} Distribution of transit depths for 143 TTV planets.
        {{\bf Bottom right:}} Distribution of predicted TESS transit depth fractional uncertainties, computed using the expressions in Appendix (A) of \citet{Price14}. The blue histogram shows the predicted depth uncertainties measured by fitting a single TESS transit. The orange and green histograms show depth uncertainties predicted by assuming all planets' during the nominal and extended mission, respectively, are perfectly phase-folded.
        }
        \label{fig:sigmatc}
\end{figure*}

In order to assess \TSS's contribution to TTV dynamical constraints, we need to estimate the precision of transit time measurements that can be derived from \TSS~light cuvres.
\cite{Price14}, building on the work of \citet{Carter08}, derive analytic formulas for the variances and covariances of transit parameters when fitting light curve photometry. They give the following expression for transit mid-time uncertainty:
\begin{equation}
  \sigma_{t_c}=\begin{cases}
    \frac{1}{Q} \sqrt{\frac{\tau T}{2}} \left(1 - \frac{\mathcal{I}}{3\tau}\right) ^ {-1/2}, & \text{if $\tau \geq \mathcal{I}$}.\\
    \frac{1}{Q} \sqrt{\frac{\mathcal{I}T}{2}} \left(1 - \frac{\tau}{3\mathcal{I}}\right) ^ {-1/2}, & \text{if $\mathcal{I} > \tau$}.
  \end{cases}
\end{equation}
\noindent{}where $T$ is the transit duration, $\tau$
is the transit ingress time,  $Q = (\delta/\sigma')\sqrt{T/\mathcal{I}}$ is the S/N for a single transit with depth $\delta$ and photometric uncertainty $\sigma'$, and $\mathcal{I}$ is the integration cadence.
We can compute the ratio of uncertainty in transit mid-time between Kepler \emph{long} cadence and \TSS~\emph{short} cadence data,
\begin{equation}
\frac{\sigma_{t_c, TESS}}{\sigma_{t_c, Kepler}} = \begin{cases}
        \frac{\sigma'_T}{\sigma'_K} \sqrt{\frac{\mathcal{I}_T}{\mathcal{I}_K}} \sqrt{\frac{1 - \tau/3\mathcal{I}_K}{1 - \tau/3\mathcal{I}_T}}, & \text{if } \mathcal{I}_T > \tau \\
        \frac{\sigma'_T}{\sigma'_K} \sqrt{\frac{\tau}{\mathcal{I}_K}} \sqrt{\frac{1 - \tau/3\mathcal{I}_K}{1 - \mathcal{I}_T/3\tau}}, & \text{if } \mathcal{I}_T \leq \tau \leq \mathcal{I}_K \\
         \frac{\sigma'_T}{\sigma'_K} \sqrt{\frac{1 - \mathcal{I}_K/3\tau}{1 - \mathcal{I}_T/3\tau}}, & \text{if } \mathcal{I}_K < \tau
    \end{cases}
    \label{eq:tc_rat}
\end{equation}
where $\mathcal{I}_T = 2$min and $\mathcal{I}_K = 30$min are the \TSS~and \KEP~cadences, respectively, and the $\sigma'$ are photometric uncertainties in a 1 hour integration.
Note that Equation \eqref{eq:tc_rat} does not dependend on transit depth or duration.

The upper left panel of Figure \ref{fig:sigmatc} shows the uncertainty ratio, Equation \eqref{eq:tc_rat}, as a function of ingress time for a representative \KEP~target star as well as the distribution of ingress times for the HL17 sample of planets. Transits with ingress duration less than 30 minutes benefit significantly from a shift to \TSS~short cadence from \KEP~long cadence, somewhat compensating for the reduced photometric performance of \TSS~relative to \KEP. The distribution of ratios of transit mid-time uncertainty, incorporating ingress time, \KEP~and \TSS~magnitudes, is in the lower panel of Figure \ref{fig:sigmatc}.

We estimate the expected \TSS~transit mid-time uncertainty for a particular planet as the product of the median mid-time uncertainty of the \citet{Rowe2015}'s \KEP~long cadence transits and the uncertainty ratio predicted by Equation \eqref{eq:tc_rat}.


\subsection{Detectability of Transits by \TSS~}
\label{s:Detect}
Our approximation for expected \TSS~transit mid-time uncertainties will fail when transits are undetectable or have very low signal-to-noise ratio (SNR).
The larger photometric uncertainties of the \TSS~mission compared to the \KEP~mission is likely to  cause a number of the planets analyzed by HL17 to be undetectable by \TSS. To illustrate this, in the bottom left panel of Figure \ref{fig:sigmatc} we plot the distribution of transit depths for the 143 TTV planets from HL17, and then in the bottom right panel of the same figure, plot the distribution of the predicted \TSS~transit depth fractional uncertainties (computed using the expressions in Appendix (A) of \citet{Price14}). The blue histogram shows the predicted depth uncertainties measured by fitting a single TESS transit.  This is a worst-case scenario for the transit depth uncertainties achievable with TESS because planets' transit SNR can be increased by phase-folding multiple transits. But phase-folding of the transits will be complicated by the planets' TTVs. However, the dynamical constraints derived from fitting the original \KEP~transit times translate to constraints on planets' possible TTV signals at the time of TESS observations. If planet masses and orbits were known exactly, systems could simply be integrated forward to the time of TESS observations and the transits could be perfectly phase-folded based on the computed transit times. This best-case scenario is represented by the orange and green histograms in Figure \ref{fig:sigmatc} which show the predicted depth uncertainties when assuming that all planetary transits during the nominal and extended mission, respectively, are perfectly phase-folded. There are 45, 81, and 121 planets to the left of the dashed line in the blue, orange, green cases, respectively: i.e. these are the number of planets that would have depths detectable at the $\sim 1\sigma$ level. 
Clearly, the true transit SNRs will lie somewhere between these idealized limits, and we anticipate that in practice, this will likely be achieved by analyzing individual systems using a detailed `photo-dynamical' modeling approach similar to that employed in, e.g., \citet{Carter11,Carter12}.

We note that previous analyses have tended to use much more stringent detection criteria when analyzing the detectability of planets in TESS data. A threshold $\sim 7\sigma$ is often used as the threshold for {\it de novo} discoveries \cite[e.g.][]{Sullivan2015}, and a $\sim 3\sigma$ threshold is often used as the threshold for recovery of known transits (such as for \KEP~planets in \TSS~data). 
We demonstrate in detail in Appendix \ref{a:Validity of PR} that a threshold $\sim1\sigma$, i.e. fractional uncertainties in transit depth $\sim1$, can yield useful data when the planet is known to transit and the period is reasonably constrained (as is the case for known \KEP~TTV systems).

\subsection{TESS Mission Scenarios} 
\label{s:Scenarios}
\begin{deluxetable}{ll lll}
\tablecaption{%
TESS Mission Scenarios. All extended scenarios are in addition to the 2-year nominal mission. Pole-Cam $C_4$ and $C_3$ indicate whether the $4^{th}$ or $3^{rd}$ cameras respectively are pointed toward the ecliptic pole. Patterns $N$ and $S$ indicate Northern or Southern hemisphere orientations respectively.
\label{t:Scen1}}
\startdata
\tablehead{
\multicolumn{2}{c}{Scenario} \\
\colhead{Symbol} &  
\colhead{Desc.} & 
\colhead{Years} & 
\colhead{Pole-Cam} & 
\colhead{Pattern}
} 
$N$             &       Nominal  &  $2$&    $C_4$ &  $SN$ \\
$E_{4,SNS}$     &       Extended &  $3$&    $C_4$ & $SNS$ \\
$E_{4,NSN}$     &       Extended &  $3$&    $C_4$ & $NSN$ \\
$E_{4,NNN}$     &       Extended &  $3$&    $C_4$ & $NNN$ \\
$E_{3,SNS}$     &       Extended &  $3$&    $C_3$ & $SNS$ \\
$E_{3,NSN}$     &       Extended &  $3$&    $C_3$ & $NSN$ \\
$E_{3,NNN}$     &       Extended &  $3$&    $C_3$ & $NNN$ \\
\enddata
\end{deluxetable}
As described in \citet{Huang18b}, the \TSS~survey divides the sky into 26 partially overlapping sectors, each of which is observed for approximately one month during the two-year nominal mission.
The first year of the mission targets the southern sky, hence the \KEP~field, which is in the north, will be covered in year two.
The \KEP~field is centered at an ecliptic latitude of $\sim65^{\circ}$: lower portions of the field will be observed for contiguous intervals of $\sim27$ days, most will receive contiguous intervals of $\sim54$ days, and some of the higher portions will be covered for $\sim78$ days during the nominal mission. We account for these differences in our simulations.

\citet{Bouma17} and \citet{Huang2018a} discuss various extended TESS mission scenarios in terms of the overall number of planet discoveries that will be expected. 
In this study, we quantify the improvements to known multi-planet TTV systems.
We consider six extended mission scenarios. 
We consider only 3-year extensions of the \TSS~mission, which we summarize in Table \ref{t:Scen1}. 
Our extended mission scenarios cover two possible camera configurations: The first, $C_4$, has camera 4 is centred on the ecliptic pole as in the nominal mission. 
The second, $C_3$, has camera 3 centered on the ecliptic pole and provides a larger area of sky with multiple pointings at the expense of coverage near the ecliptic equator. 
We also consider three possible extended mission pointing sequences: one in which \TSS~remains pointed in the northern ecliptic hemisphere for the entire extended mission ($NNN$), one in which \TSS~alternates hemispheres each year after starting in the north ($NSN$), and one in which \TSS~alternates hemispheres each year after starting in the south ($SNS$). 
As the \KEP~field is in the north, these different scenarios have the effect of adding, three, two and one extra years respectively of observations on systems in the \KEP~field.
Hence the $NNN$ scenarios ($E_{3,NNN}$ and $E_{4,NNN}$) have the greatest likelihood of improving mass measurements for TTV systems in the \KEP~field.

\section{Methods}
\label{s:Methods}

\begin{figure}[thp]
    \centering
    \includegraphics[width=\columnwidth]{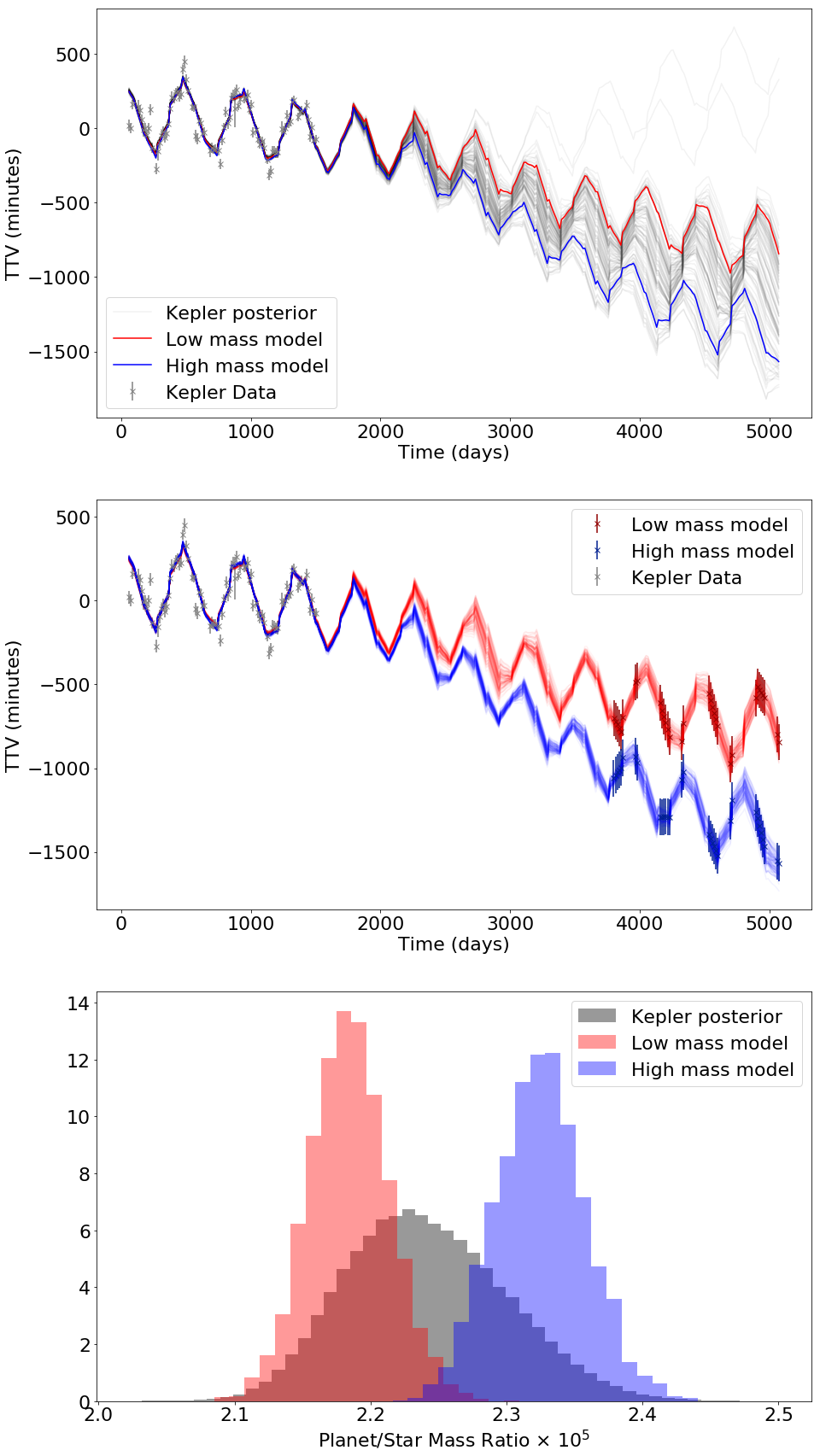}
    \caption{
    {\bf Top:} Original \KEP~TTV data points along with the posterior distribution of solutions (in gray) computed by \cite{HL17} for Kepler-36, extended out to the epoch of the \TSS~mission. 
    Two alternative ``true'' models (red and blue lines) have been selected from the posterior distribution.
    {\bf Middle:} For each ``true'' solution in the top panel, \TSS~TTVs are simulated (red and blue points) assuming a $E_{4,NNN}$ extended mission scenario, and an updated posterior distribution (red and blue swaths) calculated. 
    {\bf Bottom:} The posterior mass distributions of Kepler-36 c: the gray is for the Kepler-only data, the red and blue corresponds to the appropriate true solution illustrated in the middle panel. 
    Histogram integrals are normalized to unity. 
    The \TSS~TTV data is likely to substantially improve the precision of the mass constraints on Kepler-36 c, irrespective of whether the true model is of high or low mass.
    }
    \label{fig:K36}
\end{figure}

We now describe the method we use to estimate the likely improvement in the measured mass from \TSS~observations, illustrating our method using the Kepler-36 system. 
In the top panel of \figref{fig:K36} we illustrate the original \KEP~TTV data points for Kepler-36b, along with a sample of the posterior distribution of solutions (in gray) computed by HL17, extended out to the epoch of the \TSS~mission. 

Simulated \TSS~data were generated by randomly picking an individual from the posterior to use as an underlying true model. 
Two such examples are plotted in the top panel of \figref{fig:K36} as a thick red and blue lines.
Consider first the the red, `low mass' model: this model was integrated with TTVFast \citep{Deck14} to the end of the $E_{4,NNN}$ extended \TSS~mission,
and transits that fell within \TSS's simluated observational windows were paired with estimated uncertainties as described in Section \ref{s:Uncertainty} and plotted as red points and error-bars in the middle-panel of \figref{fig:K36}. 
We integrated the rest of the posterior samples, and computed $\chi^2$ for each set of transit times generated from the posterior sample relative to the simulated \TSS~data.

%

Next, we update the HL17 posterior samples of planet masses and orbits to reflect the new constraints of the simulated \TSS~observations.  Assuming errors in transit time measurements are Gaussian and independent, the probability of measuring a $\chi^2$-value for a series of $k$ transit times of $\chi^2(\theta)$ or greater for a particular set of planet parameters $\theta$ is given by 
\begin{equation}
    1 - F(\chi^2(\theta); k)
\end{equation}
where $F(\chi^2; k)$ is the cumulative distribution function of the chi-squared distribution with $k$ degrees of freedom. 
We therefore re-sample the HL17 posteriors, accepting each set of planet parameters $\theta$ with probability $1 - F(\chi^2(\theta); k)$.  We refer the interested reader to Appendix \ref{a:Validation of Statistical Approach} where we illustrate that the approximate approach we use and describe above is in good agreement with a more computationally demanding approach in which the entire system is re-fit to the combined constraints of \TSS~and {\it Kepler} data using Markov chain Monte Carlo. A sample of accepted solutions from the updated posterior is plotted in red in the middle panel of \figref{fig:K36}.

The ``true'' solution selected in red in \figref{fig:K36} happened to be a low mass solution, and this is reflected in the resultant mass distribution for Kepler-36 c seen in the bottom panel of \figref{fig:K36}. 
In contrast, an alternative ``true'' solution might be selected (blue lines and points in the top and middle panels of \figref{fig:K36}) in which the solution is higher mass, leading to the higher mass histogram in the bottom panel of \figref{fig:K36}. For each system in our sample, we preform 50 iterations of the posterior updating scheme described above, randomly selecting a  different ``true'' model from the HL17 posterior each time.

\subsection{Kullback-Leibler Divergence}
\label{s:KL}
To evaluate the degree of improvement represented by the updated planet mass posterior\footnote{The TTV fits constrain the planet-star mass ratios, $\mu$ and not absolute planet masses, which depends on stellar mass constraints as well as the dynamical constraints. For simplicity, however, we refer the the planet-star mass ratio posteriors as `planet mass' posteriors.}, $P$, over the HL17 planet mass posterior, $Q$, we computed the Kullback-Leibler (KL) divergence from the \KEP-derived mass distribution to the simulated \TSS~distribution. The KL divergence is defined as
\begin{equation}
    D_{KL}(P \parallel Q) = \int P(\mu) \log_2{\frac{P(\mu)}{Q(\mu)}}d\mu~
    \label{eq:KL}
\end{equation}
and quantifies the amount of information provided by the new measurements, in units of bits \citep{kullback1951}. In other words, $D_{KL}(P \parallel Q)$ is a measure of the information gained when we update our beliefs from the probability distribution $Q$ (the planet mass posterior distribution when one has only \KEP~data) to the new probability distribution $P$ (calculated with the additional information added by \TSS~data).  
A low KL divergence indicates nearly identical distributions. As an example, consider normal distributions with identical mean, i.e., $P(x) = \mathcal{N}(0, \sigma_1)$ and $Q(x) = \mathcal{N}(0, \sigma_2)$. Then, the KL divergence from $Q$ to $P$ is
\begin{equation}
    D_{KL}(P \parallel Q) = \frac{1}{2\log 2} \left( \left( \frac{\sigma_1}{\sigma_2} \right) ^2 - 1\right) - \log_2 \frac{\sigma_1}{\sigma_2}.
    \label{e:KL2}
\end{equation}
This quantity is $0$ if and only if $\sigma_1 = \sigma_2$, otherwise it is positive.

\begin{figure}[thp]
    \centering
    \includegraphics[width=0.95\columnwidth]{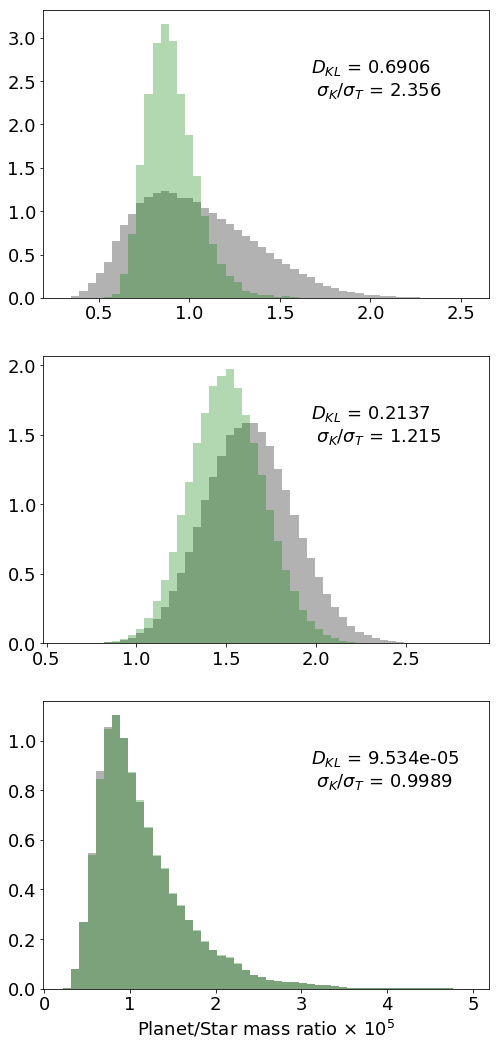}
    \caption{
    Three examples of mass-distribution changes and their corresponding KL divergences. 
    In \emph{gray} we plot the initial distributions (e.g. from Kepler-only observations).
    In \emph{green} we plot updated mass posteriors after additional observations (of the type illustrated in Figure \ref{fig:K36}). 
    The top system, with high $D_{KL}$, exhibits a significant improvement in the mass measurement.
    The middle system, exhibits a borderline significant improvement to the measured mass. 
    The bottom system, with a small $D_{L}$ value, shows negligible improvement to its measured mass. 
    We also indicate the ratio of the standard deviations of the initial ($\sigma_{Mass,K}$) and updated ($\sigma_{Mass,T}$) distributions.
    }
    \label{fig:K36b}
\end{figure}

To estimate the KL divergence between two mass posteriors, we compute a kernel density estimation (KDE) of each posterior and then approximate the integral as a Riemann sum over the range of masses, according to Eq. \ref{eq:KL}.

To provide some intuition for the improvement in the measured mass that corresponds to a given KL divergence, we plot in \figref{fig:K36b} some examples of systems (deliberately anonymized) with high, medium and low KL divergences.
Here we have plotted the \KEP~mass posterior ($Q$) in gray and the updated posterior ($P$) in green.

\section{Results: Expected Improvements to Mass Constraints}
\label{s:Nominal}
\begin{figure*}[thp]
    \centering
    \includegraphics[width=0.95\textwidth]{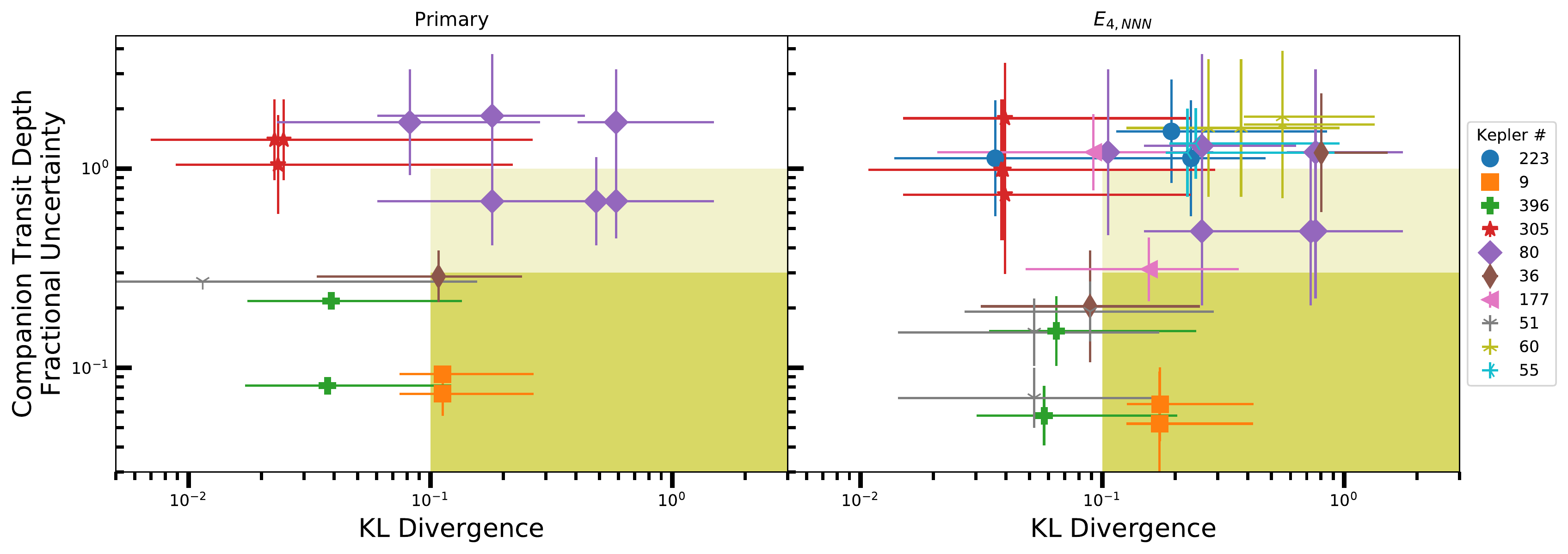}
    \caption{%
    Planetary K-L divergence versus companion transit depth uncertainty.
    {\bf Left:}
    Results for the \emph{nominal} mission.
    {\bf Right:}
    Results for the \emph{$E_{4,NNN}$ extended} mission.
    The vertical bars span the most pessimistic depth uncertainty (top-of-bar) when assuming single-transit-only measurements, to the most optimistic (bottom of bar) when all transits are perfectly stacked. 
    The horizontal bars indicate the central $68\%$ of the spread of the KL divergences for each simulation.
    We plot only the \NPE{} planets in the \NSE{} systems whose median KL divergence and companion transit depth uncertainty comes within the larger shaded box (in either the extended or nominal missions): these systems are both easily detected by \TSS, and would have the greatest improvement to their mass measurements.
    }
    \label{fig:kl_delta}
\end{figure*}
We analyze each of the mission scenarios (nominal and extended) described in Section \ref{s:Scenarios}. 
For each system under each of the mission scenarios considered, we uniformly selected 50 different alternative true models from the \KEP~posterior and computed an updated posterior with the procedure described in Section \ref{s:Methods}. 

In Figure \ref{fig:kl_delta} we plot the planetary K-L divergence against the transit depth uncertainty of a \emph{companion} planet.\footnote{We plot the depth of the companion planet, not the target planet, because it is a detection of TTVs in the companion that allows constraints on the mass of the target planet. For planets having more than one transiting companion, multiple points are plotted with the same KL divergence at each of the different fractional depth uncertainties of the associated companions.} The results for the \emph{nominal} mission are in the left-hand panel, and then as a comparison the results for the $E_{4,NNN}$ extended mission are provided in the right-hand panel (while we analyzed all extended mission scenarios, we present only this most optimistic scenario with three additional years of coverage.). 

The range indicated by the vertical bars illustrates the most pessimistic depth uncertainty (top of bar) when assuming single-transit-only measurements, to the most optimistic depth uncertainty (bottom of bar) for perfect stacking of all detectable transits. 
The horizontal bars indicate the central $68\%$ of the spread of the KL divergences for each simulation.

As discussed in Section \ref{s:Detect}, a fractional transit depth uncertainty $\sim1.0$ is likely marginally detectable, given a known transiting planet.
In Appendix \ref{a:Validity of PR} (Figure \ref{f:depth}), we explicitly demonstrate that meaningful measurements of transit mid-times can be achieved for \emph{many} systems with fractional transit depth uncertainties $\sim1.0$.
We over-plot in Figure \ref{fig:kl_delta} a shaded box to indicate the region of parameter-space that would have both significantly improved mass measurements (KL divergence $\gsim0.1$) as well as being plausibly detectable (fractional transit depth uncertainty $\sigma_\delta/\delta\lsim1.0$).  
We plot \emph{only} the planets whose ``error bars'' touch the shaded box (omitted systems would likely be undetectable and/or have negligible improvements to their measured masses).
We also over-plot (darker shaded box) a more conservative detection threshold with a fractional transit depth uncertainty $\sigma_\delta/\delta\lsim0.3$.

There are \NPN{} planets across \NSN{} systems that fall within the larger shaded box highlighted in Figure \ref{fig:kl_delta} in the nominal mission, and \NPE{} planets across \NSE{} systems in the extended, $E_{4,NNN}$, mission: these systems are both easily detected by \TSS, and would have the greatest improvement to their mass measurements. For the more conservative restriction of $\sigma_\delta/\delta<0.3$, these numbers fall to \NPNCons{} planets across \NSNCons{} systems in the nominal mission, and \NPECons{} planets across \NSECons{} systems in the extended, $E_{4,NNN}$, mission.

Thus far, we have presented the results for the nominal mission and for the extended mission scenario $E_{4,NNN}$. 
As expected, when we analyze in detail the alternative extended mission scenarios in Table \ref{t:Scen1} that contain fewer observation years in the northern hemisphere, we find that the number of planets with improved mass measurements is intermediate between the results of the nominal and $E_{4,NNN}$ scenarios presented above. We do not find altering the pole-centered camera from $4$ to $3$ has a significant impact on our results.

\begin{figure}[thp]
    \centering
    \includegraphics[width=0.95\columnwidth]{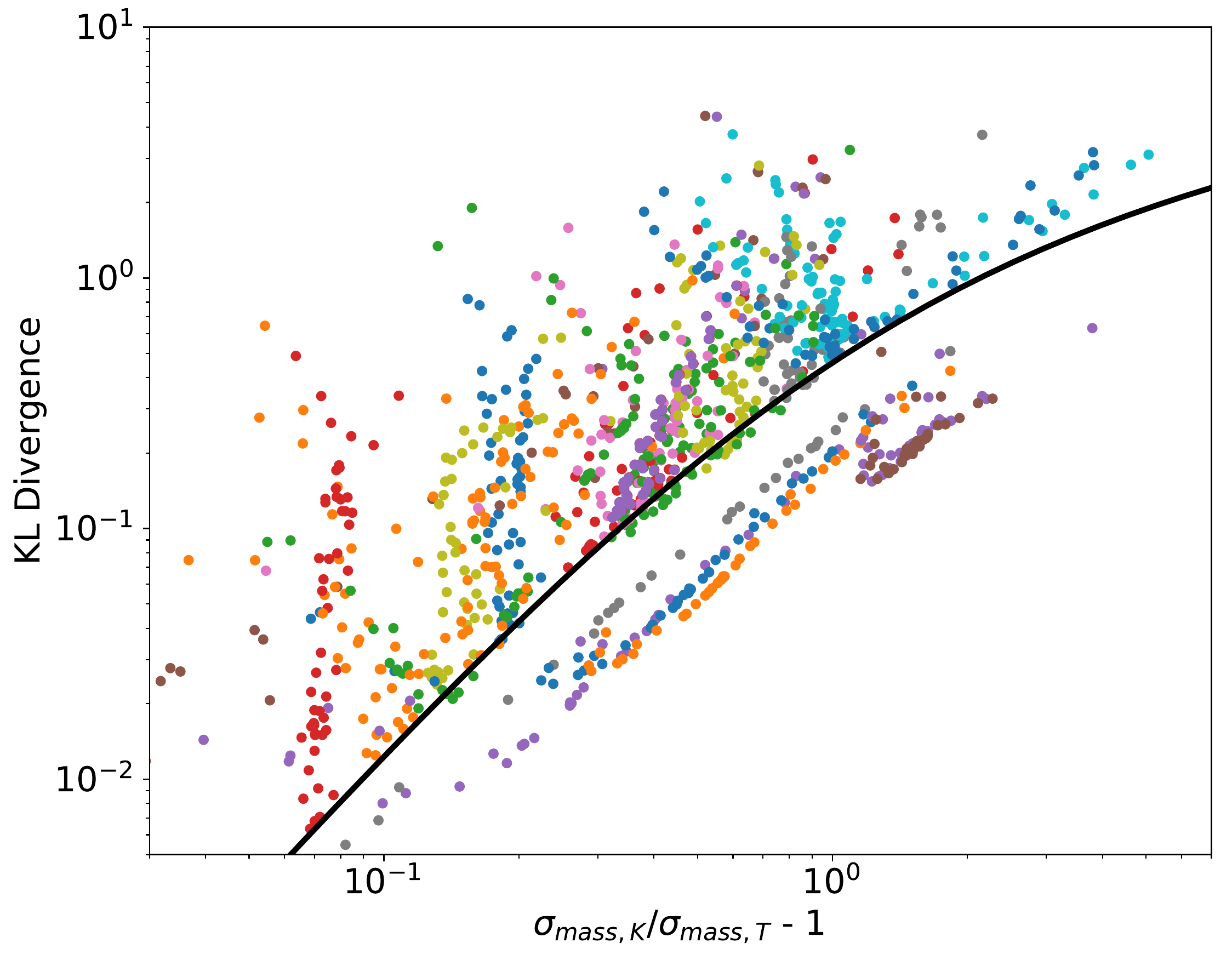}
    \caption{ We plot $\sigma_{Mass,K}/\sigma_{Mass,T}$ (the same quantity indicated in the labels of Figure \ref{fig:K36b}) against the KL divergence for the 50 true models evaluated for the extended $E_{4,NNN}$ scenario, showing the results for each of the \NPE{} planets from Figure \ref{fig:kl_delta}.
    Results for each planet use a common color.
    For comparison, the solid black curve shows the expected results for Gaussian distributions assuming the same mean for prior and posterior, given by Equation \ref{e:KL2}.
    }
    \label{fig:scatter}
\end{figure}

Given both the unfamiliarity of the KL divergence statistic, as well as the significant non-Gaussianities in the posterior distributions, in Figure \ref{fig:scatter} we plot the underlying quantity $\sigma_{Mass,K}/\sigma_{Mass,T}$ (i.e. the same quantity indicated in the labels of Figure \ref{fig:K36b}) against the KL divergence from all 50 simulation iterations of the \NPE{} planets in the right-hand panel of Figure \ref{fig:kl_delta}, showing the results for each planet using a different color. 
We over plot (solid line) the expected results for a Gaussian distribution that has the same mean in the prior and posterior distributions. 
The scatter of points for each planet reflects the significant non-Gaussianity present in the distributions, as well as clarifying the overall scale of the variation seen in the mass-measurement improvements (for a single planet) across the 50 different simulations performed for each.

\begin{figure*}[thp]
    \centering
    \includegraphics[trim = 100mm 75mm 100mm 75mm, clip, angle=0, width=\textwidth]{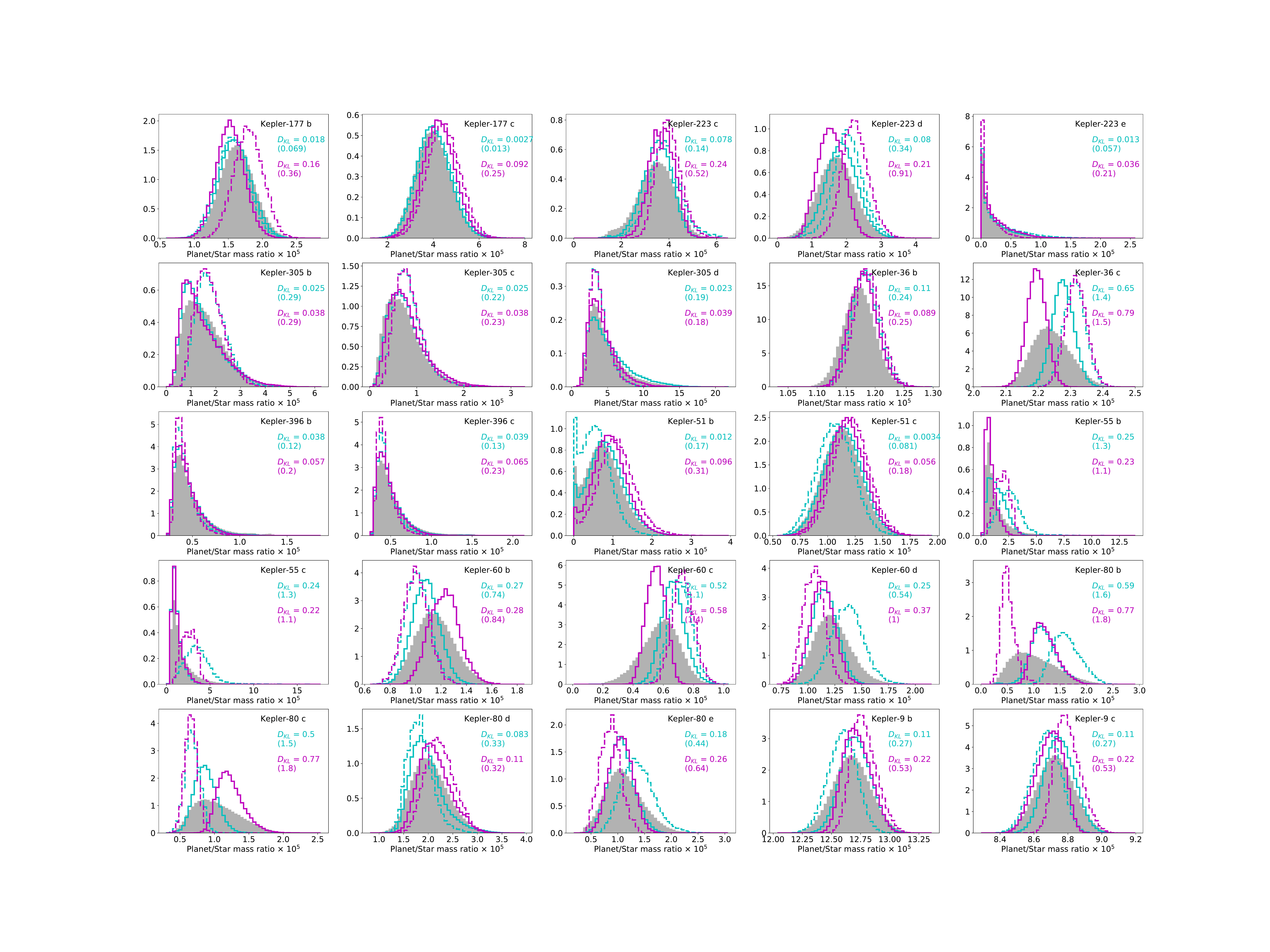}
    \caption{Mass histograms of the \NPE{} planets that are most-improved through \TSS~observations (i.e., those planets from Figure \ref{fig:kl_delta} ). 
    Gray bars are the original \KEP~posterior, cyan is after a simulated \TSS~primary mission, and magenta after the $E_{4,NNN}$ extended mission.
    Solid lines are the median results, dashed lines correspond to the upper-limits of the horizontal error bars in Figure 
    \ref{fig:kl_delta}.
    Values for the KL divergence are given in each panel, with the values for the dashed upper-limit histograms being in parentheses.  
}
\label{fig:portraits}
\end{figure*}

To make explicitly clear which individual planets and systems  we expect to benefit the most from additional \TSS~data, we plot in Figure \ref{fig:portraits} for all \NPE{} planets their mass histograms from the \KEP~mission (gray), the nominal \TSS~mission (cyan) and the $E_{4,NNN}$ extended \TSS~mission (magenta). 
We use solid lines for histograms representing the median KL divergence case, and dashed lines for the histograms corresponding to the upper-limits of the horizontal error bars in Figure \ref{fig:kl_delta}.
We see that planets such as Kepler 80-b and Kepler 80-c will benefit significantly from the acquisition of \TSS~short cadence data.

\section{Discussion}
\label{SECN:DISC}
The main goal of this work has been to provide a quantitative estimate of the potential for the Transiting Exoplanet Survey Satellite (\TSS) to improve the mass (and eccentricity) constraints obtained from transit timing variations (TTVs) observed in \KEP~systems. 

We find that:
\begin{enumerate}
\item{
Approximately \NPNRange{} planets will have their measured masses (and eccentricities) significantly improved (i.e. their KL-divergence is $\gsim 0.1$) by measurements taken during the nominal \TSS~mission;
}
\item{ 
Approximately \NPERange{} planets (total) will have their measured masses and eccentricities significantly improved through measurements taken during a nominal$+$extended \TSS~mission;
}
\end{enumerate}

We note that we have not evaluated other benefits that may flow from the observations of these (and other) systems at short cadence. 
In particular, the acquisition of such data may enable the detection (or constraint) of orbit evolution effects such as semi-major axis evolution driven by tidal effects.
We refer the interested reader to the work of Christ et al. (\emph{in prep.}) for a detailed examination of such issues.

\facilities{Exoplanet Archive}
\software{TTVFast, Matplotlib}

\section{Acknowledgments}


MJH and MJP gratefully acknowledge 
NASA Origins of Solar Systems Program grant \\ NNX13A124G, 
NASA Origins of Solar Systems Program grant NNX10AH40G via sub-award agreement 1312645088477, 
NASA Solar System Observations grant NNX16AD69G, 
as well as support from the Smithsonian 2015 CGPS/Pell Grant program.
MG gratefully acknowledges the Origins of Life Summer Undergraduate Research Prize Award Program.
SH gratefully acknowledges the CfA Fellowship.
The computations in this paper were run on the Odyssey cluster supported by the FAS Science Division Research Computing Group at Harvard University.
This research has made use of the NASA Exoplanet Archive, which is operated by the California Institute of Technology, under contract with the National Aeronautics and Space Administration under the Exoplanet Exploration Program


\appendix

\section{Validity of Transit Mid-Time Approximations}
\label{a:Validity of PR}
\begin{figure}[thp]
    \centering
    \includegraphics[width=0.95\textwidth]{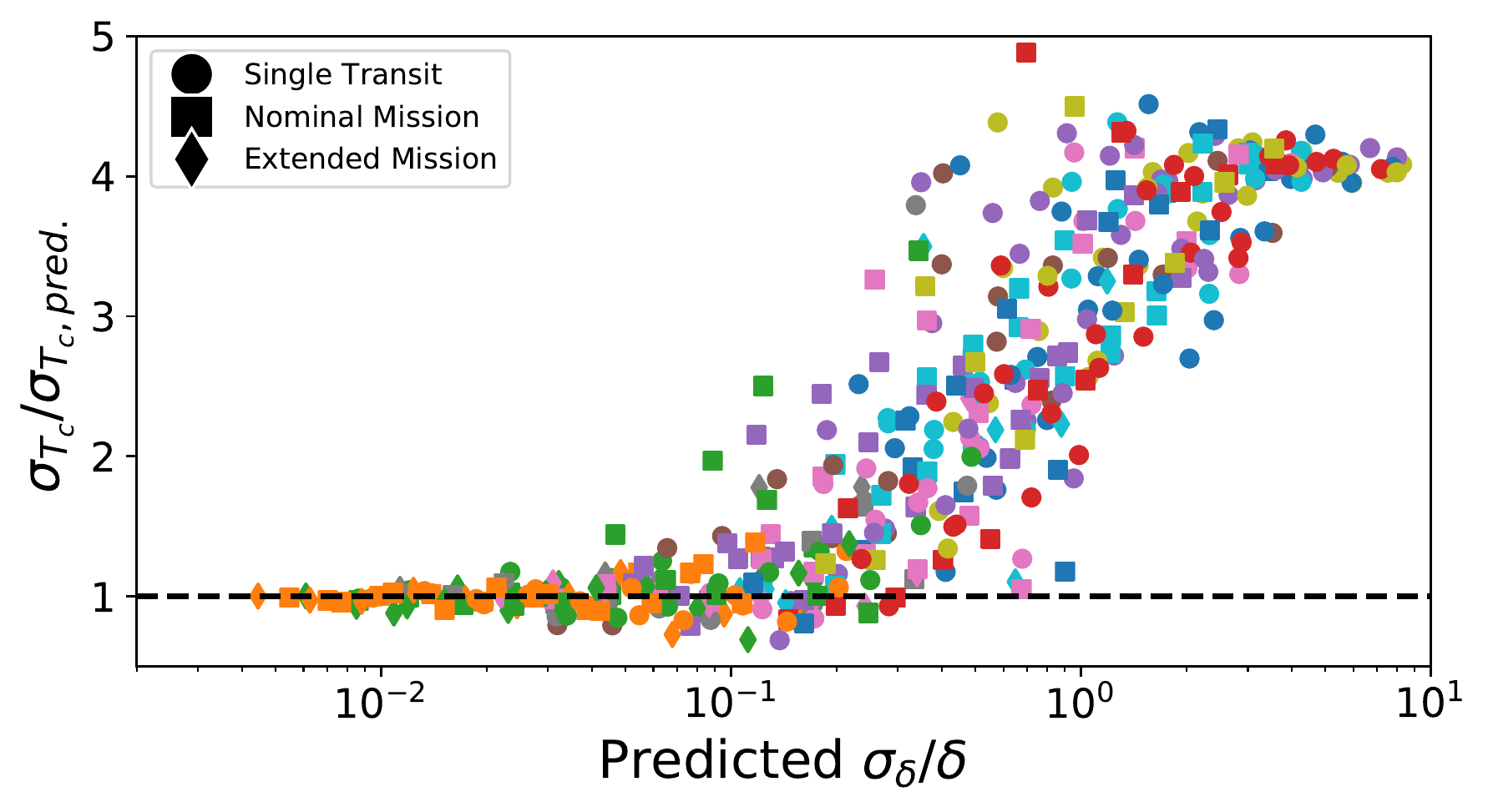}
    \caption{Transit mid-time uncertainties, $\sigma_{T_c}$, recovered from fitting simulated light-curves and normalized by $\sigma_{T_c,pred.}$, the uncertainty predicted by \citet{Price14}'s formulas derived from a Fisher information analysis versus the predicted fractional depth uncertainty, $\sigma_\delta/\delta$, also computed from \citet{Price14}'s formulas. Each point is computed by fitting a synthetic light curve generated by adopting the transit properties of one of the {\it Kepler} planets plotted in Figure \ref{fig:kl_delta} and a photometric noise level $\sigma_{phot}\in[0.5\sigma_{phot,Kep},1.5\sigma_{phot,TESS}]$.
    Symbol colors denote the {\it Kepler} system as in Figure \ref{fig:kl_delta}. Different symbols are used for points computed using sythetic light curves corresponding to either a single transit (circles) or phase-folded data containing the number of transits expected during the nominal (square) or extended (diamond) missions.}
    \label{f:depth}
\end{figure}

To understand whether the relation in Eq. \ref{eq:tc_rat} for the uncertainty in the mid-time of the transits holds for the low signal-to-noise ratio transits observed by \TSS, we created synthetic light curves of the 25 planets plotted in Figure \ref{fig:kl_delta} using the \texttt{batman} python module \citep{BATMAN}.
The light curves are sampled at a 2 minute cadence and the transit properties are based on the planet periods, planet-star radius ratios, impact parameters, transit durations, and ingress times reported on Exoplanet Archive. For simplicity, we ignore limb-darkening.
We generate synthetic light curves with mid-transit time $T_c=0$ by adding random Gaussian noise for 10 different levels of photometric noise logarithmically spaced between  $\sigma_{phot}\in[0.5\sigma_{phot,Kep},1.5\sigma_{phot,TESS}]$ where $\sigma_{phot,Kep}$ and $\sigma_{phot,TESS}$ are the photmetric precision of {\it Kepler} and \TSS, respectively.
We then attempt to recover the transit mid-time from each synthetic light curve by fitting the trapezoidal transit model presented in \citet{Price14}.
We fit the transit mid-time by computing $\chi^2(T_c)$ over a grid of values spanning $\pm 7\sigma_{T_c,pred.}$ where $\sigma_{T_c,pred.}$ is the predicted transit mid-time uncertainty based on \citet{Price14}'s formula. All transit parameters other than the mid-time are fixed. We construct a posterior distribution, $p(T_c)$, for the transit mid-time from the grid of $\chi^2$ values such that $p(T_c)\propto \exp[-\frac{1}{2}\chi^2(T_c)]$ and then compute the standard deviation,$\sigma_{T_c}$ of this posterior distribution in order to compare it with the prediction,  $\sigma_{T_c,pred.}$, of \citet{Price14}'s formula. 
This procedure is repeated three times for each planet: first, for light-curve data containing a single transit and then twice more for phase-folded data containing the number of transits expected during TESS's nominal mission and extended mission $E_{4,NNN}$. The number of nominal and extended mission transits are taken to be $54\text{~d.}/P$ and  $216\text{~d.}/P$, respectively, where $P$ is the planet's orbital period.

Results are shown in Figure \ref{f:depth}, where computed values of $\sigma_{T_c}/\sigma_{T_c,pred.}$ are plotted  against the expected transit depth fractional uncertainty, $\sigma_\delta/\delta$, predicted by \citet{Price14}'s analysis, which depends on the level of photometric noise as well as the particular properties of the transit. From the Figure, we see that \citet{Price14}'s formula under-predicts transit mid-time uncertainties for $\sigma_\delta/\delta \gtrsim 0.3$.

\section{Validation of Rejection-Sampling Approach}
\label{a:Validation of Statistical Approach}
\begin{figure}[thp]
    \centering
    \includegraphics[width=0.95\columnwidth]{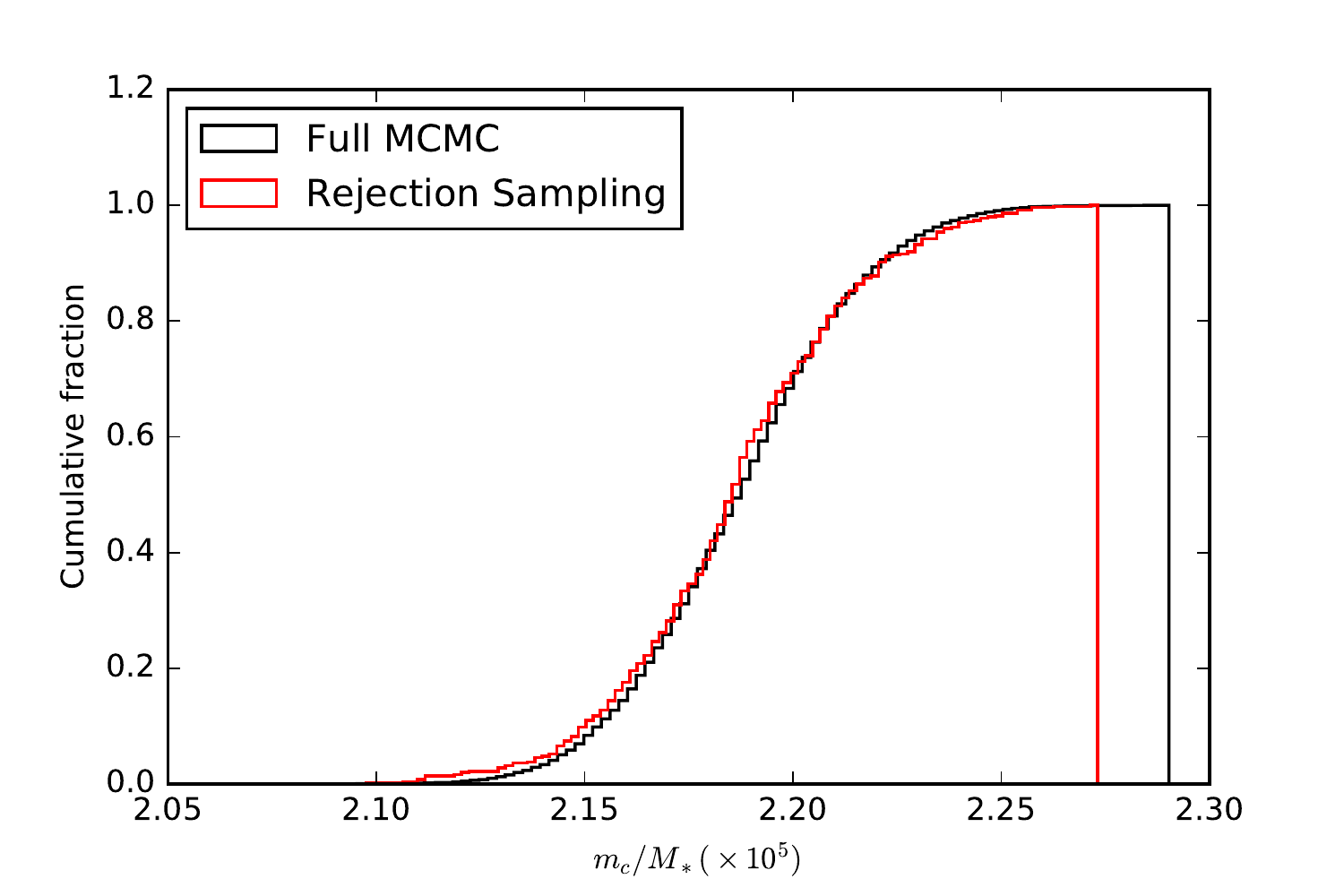}
    \caption{ Comparison between the full MCMC modelling approach of the kind employed by HL17 (black) and the rejection sampling method (red) employed in the main body of this current paper.
    We find that the results are essentially indistinguishable, providing confidence that our rejection sampling method is robust. 
    }
    \label{f:comparison}
\end{figure}
In Section \ref{s:Methods} we described the rejection sampling method we employed to determine the statistical power of the adding additional observations.
One could employ a more computationally intensive approach to this problem, and instead of using rejection sampling, one can undertake the full MCMC modelling approach employed by HL17 for the Kepler data-set, and undertake a complete re-fitting of the data using the combined Kepler + simulated-\TSS~data-set. 

Here we provide a comparison of the Kepler-36 results obtained using the full MCMC remodelling method, with the rejection sampling method described in Section \ref{s:Methods}. We plot the results in Figure \ref{f:comparison}, comparing the CDF histograms for Kepler-36c's mass measured with the two methods.  We find that the results are essentially indistinguishable, providing confidence that our rejection sampling method is robust. 

\begin{thebibliography}{}
\expandafter\ifx\csname natexlab\endcsname\relax\def\natexlab#1{#1}\fi
\providecommand{\url}[1]{\href{#1}{#1}}

\bibitem[{{Agol} {et~al.}(2005){Agol}, {Steffen}, {Sari}, \&
  {Clarkson}}]{Agol05}
{Agol}, E., {Steffen}, J., {Sari}, R., \& {Clarkson}, W. 2005, \mnras, 359, 567

\bibitem[{Barclay(2017)}]{ticgen}
Barclay, T. 2017, tessgi/ticgen: v1.0.0, , , doi:10.5281/zenodo.888217.
\newblock \url{https://doi.org/10.5281/zenodo.888217}

\bibitem[{{Borucki} {et~al.}(2010){Borucki}, {Koch}, {Basri}, {Batalha},
  {Brown}, {Caldwell}, {Caldwell}, {Christensen-Dalsgaard}, {Cochran},
  {DeVore}, {Dunham}, {Dupree}, {Gautier}, {Geary}, {Gilliland}, {Gould},
  {Howell}, {Jenkins}, {Kondo}, {Latham}, {Marcy}, {Meibom}, {Kjeldsen},
  {Lissauer}, {Monet}, {Morrison}, {Sasselov}, {Tarter}, {Boss}, {Brownlee},
  {Owen}, {Buzasi}, {Charbonneau}, {Doyle}, {Fortney}, {Ford}, {Holman},
  {Seager}, {Steffen}, {Welsh}, {Rowe}, {Anderson}, {Buchhave}, {Ciardi},
  {Walkowicz}, {Sherry}, {Horch}, {Isaacson}, {Everett}, {Fischer}, {Torres},
  {Johnson}, {Endl}, {MacQueen}, {Bryson}, {Dotson}, {Haas}, {Kolodziejczak},
  {Van Cleve}, {Chandrasekaran}, {Twicken}, {Quintana}, {Clarke}, {Allen},
  {Li}, {Wu}, {Tenenbaum}, {Verner}, {Bruhweiler}, {Barnes}, \&
  {Prsa}}]{Borucki2010}
{Borucki}, W.~J., {Koch}, D., {Basri}, G., {et~al.} 2010, Science, 327, 977

\bibitem[{{Bouma} {et~al.}(2017){Bouma}, {Winn}, {Kosiarek}, \&
  {McCullough}}]{Bouma17}
{Bouma}, L.~G., {Winn}, J.~N., {Kosiarek}, J., \& {McCullough}, P.~R. 2017,
  ArXiv e-prints, arXiv:1705.08891

\bibitem[{{Carter} {et~al.}(2008){Carter}, {Yee}, {Eastman}, {Gaudi}, \&
  {Winn}}]{Carter08}
{Carter}, J.~A., {Yee}, J.~C., {Eastman}, J., {Gaudi}, B.~S., \& {Winn}, J.~N.
  2008, \apj, 689, 499

\bibitem[{{Carter} {et~al.}(2011){Carter}, {Fabrycky}, {Ragozzine}, {Holman},
  {Quinn}, {Latham}, {Buchhave}, {Van Cleve}, {Cochran}, {Cote}, {Endl},
  {Ford}, {Haas}, {Jenkins}, {Koch}, {Li}, {Lissauer}, {MacQueen}, {Middour},
  {Orosz}, {Rowe}, {Steffen}, \& {Welsh}}]{Carter11}
{Carter}, J.~A., {Fabrycky}, D.~C., {Ragozzine}, D., {et~al.} 2011, Science,
  331, 562

\bibitem[{{Carter} {et~al.}(2012){Carter}, {Agol}, {Chaplin}, {Basu},
  {Bedding}, {Buchhave}, {Christensen-Dalsgaard}, {Deck}, {Elsworth},
  {Fabrycky}, {Ford}, {Fortney}, {Hale}, {Handberg}, {Hekker}, {Holman},
  {Huber}, {Karoff}, {Kawaler}, {Kjeldsen}, {Lissauer}, {Lopez}, {Lund},
  {Lundkvist}, {Metcalfe}, {Miglio}, {Rogers}, {Stello}, {Borucki}, {Bryson},
  {Christiansen}, {Cochran}, {Geary}, {Gilliland}, {Haas}, {Hall}, {Howard},
  {Jenkins}, {Klaus}, {Koch}, {Latham}, {MacQueen}, {Sasselov}, {Steffen},
  {Twicken}, \& {Winn}}]{Carter12}
{Carter}, J.~A., {Agol}, E., {Chaplin}, W.~J., {et~al.} 2012, Science, 337, 556

\bibitem[{{Deck} {et~al.}(2014){Deck}, {Agol}, {Holman}, \&
  {Nesvorn{\'y}}}]{Deck14}
{Deck}, K.~M., {Agol}, E., {Holman}, M.~J., \& {Nesvorn{\'y}}, D. 2014, \apj,
  787, 132

\bibitem[{{Hadden} \& {Lithwick}(2017)}]{HL17}
{Hadden}, S., \& {Lithwick}, Y. 2017, \aj, 154, 5

\bibitem[{{Holczer} {et~al.}(2016){Holczer}, {Mazeh}, {Nachmani},
  {Jontof-Hutter}, {Ford}, {Fabrycky}, {Ragozzine}, {Kane}, \&
  {Steffen}}]{Holczer2016}
{Holczer}, T., {Mazeh}, T., {Nachmani}, G., {et~al.} 2016, \apjs, 225, 9

\bibitem[{{Holman} \& {Murray}(2005)}]{Holman05}
{Holman}, M.~J., \& {Murray}, N.~W. 2005, Science, 307, 1288

\bibitem[{{Holman} {et~al.}(2010){Holman}, {Fabrycky}, {Ragozzine}, {Ford},
  {Steffen}, {Welsh}, {Lissauer}, {Latham}, {Marcy}, {Walkowicz}, {Batalha},
  {Jenkins}, {Rowe}, {Cochran}, {Fressin}, {Torres}, {Buchhave}, {Sasselov},
  {Borucki}, {Koch}, {Basri}, {Brown}, {Caldwell}, {Charbonneau}, {Dunham},
  {Gautier}, {Geary}, {Gilliland}, {Haas}, {Howell}, {Ciardi}, {Endl},
  {Fischer}, {F{\"u}r{\'e}sz}, {Hartman}, {Isaacson}, {Johnson}, {MacQueen},
  {Moorhead}, {Morehead}, \& {Orosz}}]{Holman10}
{Holman}, M.~J., {Fabrycky}, D.~C., {Ragozzine}, D., {et~al.} 2010, Science,
  330, 51

\bibitem[{{Huang} {et~al.}(2018{\natexlab{a}}){Huang}, {Burt}, {Vanderburg},
  {G{\"u}nther}, {Shporer}, {Dittmann}, {Winn}, {Wittenmyer}, {Sha}, {Kane},
  {Ricker}, {Vanderspek}, {Latham}, {Seager}, {Jenkins}, {Caldwell}, {Collins},
  {Guerrero}, {Smith}, {Quinn}, {Udry}, {Pepe}, {Bouchy}, {gransan}, {Lovis},
  {Ehrenreich}, {Marmier}, {Mayor}, {Wohler}, {Haworth}, {Morgan}, {Fausnaugh},
  {Charbonneau}, {Narita}, \& {the TESS team}}]{Huang18b}
{Huang}, C.~X., {Burt}, J., {Vanderburg}, A., {et~al.} 2018{\natexlab{a}},
  ArXiv e-prints, arXiv:1809.05967

\bibitem[{{Huang} {et~al.}(2018{\natexlab{b}}){Huang}, {Shporer}, {Dragomir},
  {Fausnaugh}, {Levine}, {Morgan}, {Nguyen}, {Ricker}, {Wall}, {Woods}, \&
  {Vanderspek}}]{Huang2018a}
{Huang}, C.~X., {Shporer}, A., {Dragomir}, D., {et~al.} 2018{\natexlab{b}},
  ArXiv e-prints, arXiv:1807.11129

\bibitem[{{Jontof-Hutter} {et~al.}(2016){Jontof-Hutter}, {Ford}, {Rowe},
  {Lissauer}, {Fabrycky}, {Van Laerhoven}, {Agol}, {Deck}, {Holczer}, \&
  {Mazeh}}]{Jontof-Hutter2016}
{Jontof-Hutter}, D., {Ford}, E.~B., {Rowe}, J.~F., {et~al.} 2016, \apj, 820, 39

\bibitem[{Kreidberg(2015)}]{BATMAN}
Kreidberg, L. 2015, Publications of the Astronomical Society of the Pacific,
  127, 1161.
\newblock \url{http://stacks.iop.org/1538-3873/127/i=957/a=1161}

\bibitem[{Kullback \& Leibler(1951)}]{kullback1951}
Kullback, S., \& Leibler, R.~A. 1951, Ann. Math. Statist., 22, 79.
\newblock \url{https://doi.org/10.1214/aoms/1177729694}

\bibitem[{{Lithwick} {et~al.}(2012){Lithwick}, {Xie}, \& {Wu}}]{Lithwick12}
{Lithwick}, Y., {Xie}, J., \& {Wu}, Y. 2012, \apj, 761, 122

\bibitem[{{Ofir} {et~al.}(2018){Ofir}, {Xie}, {Jiang}, {Sari}, \&
  {Aharonson}}]{Ofir2018}
{Ofir}, A., {Xie}, J.-W., {Jiang}, C.-F., {Sari}, R., \& {Aharonson}, O. 2018,
  \apjs, 234, 9

\bibitem[{{Price} \& {Rogers}(2014)}]{Price14}
{Price}, E.~M., \& {Rogers}, L.~A. 2014, \apj, 794, 92

\bibitem[{{Ricker} {et~al.}(2015){Ricker}, {Winn}, {Vanderspek}, {Latham},
  {Bakos}, {Bean}, {Berta-Thompson}, {Brown}, {Buchhave}, {Butler}, {Butler},
  {Chaplin}, {Charbonneau}, {Christensen-Dalsgaard}, {Clampin}, {Deming},
  {Doty}, {De Lee}, {Dressing}, {Dunham}, {Endl}, {Fressin}, {Ge}, {Henning},
  {Holman}, {Howard}, {Ida}, {Jenkins}, {Jernigan}, {Johnson}, {Kaltenegger},
  {Kawai}, {Kjeldsen}, {Laughlin}, {Levine}, {Lin}, {Lissauer}, {MacQueen},
  {Marcy}, {McCullough}, {Morton}, {Narita}, {Paegert}, {Palle}, {Pepe},
  {Pepper}, {Quirrenbach}, {Rinehart}, {Sasselov}, {Sato}, {Seager},
  {Sozzetti}, {Stassun}, {Sullivan}, {Szentgyorgyi}, {Torres}, {Udry}, \&
  {Villasenor}}]{TESS15}
{Ricker}, G.~R., {Winn}, J.~N., {Vanderspek}, R., {et~al.} 2015, Journal of
  Astronomical Telescopes, Instruments, and Systems, 1, 014003

\bibitem[{{Rowe} {et~al.}(2015){Rowe}, {Coughlin}, {Antoci}, {Barclay},
  {Batalha}, {Borucki}, {Burke}, {Bryson}, {Caldwell}, {Campbell},
  {Catanzarite}, {Christiansen}, {Cochran}, {Gilliland}, {Girouard}, {Haas},
  {He{\l}miniak}, {Henze}, {Hoffman}, {Howell}, {Huber}, {Hunter},
  {Jang-Condell}, {Jenkins}, {Klaus}, {Latham}, {Li}, {Lissauer}, {McCauliff},
  {Morris}, {Mullally}, {Ofir}, {Quarles}, {Quintana}, {Sabale}, {Seader},
  {Shporer}, {Smith}, {Steffen}, {Still}, {Tenenbaum}, {Thompson}, {Twicken},
  {Van Laerhoven}, {Wolfgang}, \& {Zamudio}}]{Rowe2015}
{Rowe}, J.~F., {Coughlin}, J.~L., {Antoci}, V., {et~al.} 2015, \apjs, 217, 16

\bibitem[{{Stassun} {et~al.}(2017){Stassun}, {Oelkers}, {Pepper}, {Paegert},
  {DeLee}, {Torres}, {Latham}, {Charpinet}, {Dressing}, {Huber}, {Kane},
  {Lepine}, {Mann}, {Muirhead}, {Rojas-Ayala}, {Silvotti}, {Fleming}, {Levine},
  {Plavchan}, \& {the TESS Target Selection Working Group}}]{TIC2017}
{Stassun}, K.~G., {Oelkers}, R.~J., {Pepper}, J., {et~al.} 2017, ArXiv
  e-prints, arXiv:1706.00495

\bibitem[{{Sullivan} {et~al.}(2015){Sullivan}, {Winn}, {Berta-Thompson},
  {Charbonneau}, {Deming}, {Dressing}, {Latham}, {Levine}, {McCullough},
  {Morton}, {Ricker}, {Vanderspek}, \& {Woods}}]{Sullivan2015}
{Sullivan}, P.~W., {Winn}, J.~N., {Berta-Thompson}, Z.~K., {et~al.} 2015, \apj,
  809, 77

\end{thebibliography}
\end{document}